# Closing the Information Gap in Unidentified Anomalous Phenomena (UAP) Studies

Gretchen R. Stahlman[1][0000-0001-8814-863X]

[1] Florida State University, Tallahassee FL 32306, USA
gstahlman@fsu.edu

**Abstract.** Unidentified Anomalous Phenomena (UAP), also known as Unidentified Flying Objects (UFOs), has shifted from being a stigmatized topic on the fringes of scientific inquiry to a legitimate subject of scientific interest with a need for high quality, curated data, and rigorous scientific investigation. This paper presents a preliminary scoping review and analysis of scholarly literature related to UAP from 1967 until 2023, exploring a diverse range of research areas across disciplines to illustrate scholarly discourse about the topic. The paper focuses on characterizing papers published in recent years and notes that Library & Information Science is unrepresented in the current UAP literature. The paper also discusses how researchers across the iFields can contribute to UAP studies through inherent expertise such as data curation and data science as well as information behavior and information literacy, among others. The paper concludes by emphasizing that UAP Studies offer a rich intellectual realm for information science research, with the iFields well positioned to play a crucial role in supporting and engaging in the study of UAP.

## 1   Background

Unidentified Anomalous Phenomena (UAP), also known as Unidentified Flying Objects (UFOs), have long captured the public's fascination and speculation while stigmatized in scientific circles [1]. The notion of extraterrestrial UFOs visiting Earth has been prevalent in popular culture for the better part of a century. However, the recently re-branded concept "UAP" emerged to represent an increasingly legitimate area of scientific interest and inquiry [2, 3], relatively agnostic about the "extraterrestrial" hypothesis in pursuit of rational explanations. Current UAP research encompasses a variety of scientific domains, meanwhile lacking in rigorous methods and sources of reliable data.

   This shift in perception gained traction with a 2021 public report released by the U.S. Office of the Director of National Intelligence (ODNI) [4]. The report acknowledged that some incidents of strange objects in the sky remain unexplained and that more and better-quality data are needed to understand the nature of these objects, emphasizing a need for further investigation about UAP. The report therefore sparked renewed scientific interest and public discourse and led to the formation of task forces, agencies, and initiatives to investigate UAP.

One such task force, NASA's Unidentified Anomalous Phenomena Independent Study Team (UAPIST), recently worked to identify existing data and to make recommendations for future data collection efforts to support scientific study of UAP [5]. The task force aimed to establish a roadmap for answering questions such as, "Are these objects real or are they sensor artifacts? Are they a threat to aerospace safety? Are they a threat to U.S. national security? Are they unknown natural phenomena? What else could they be?" [6]. While the study team reports no conclusive evidence of an extraterrestrial origin for UAP, it urges a continued role for NASA in U.S.-government efforts to study UAP. Recommendations primarily focus on a critical need for de-stigmatizing the topic to encourage reporting and research, as well as high-quality, curated data adhering to the FAIR principles for Findability, Accessibility, Interoperability, and Reusability [7]. Ultimately, the NASA UAPIST report and other recent developments emphasize that understanding UAP is a data curation and data science problem, with human and social science implications as well.

The broader search for evidence of life beyond Earth has catalyzed scientific progress in astronomy and related fields such as planetary science, exobiology, and the Search for Extraterrestrial Intelligence (SETI) [8]. However, scientific study of UAP is subject to unique challenges and may be especially vulnerable to misinterpretation and sensationalism [9, 10], indicating a timely need for strategic action among the iFields (i.e., information studies, broadly conceived, as defined in [11]). This paper presents an exploratory study on existing research and opportunities for future work in UAP Studies. A literature review was conducted targeting UAP-related publications, with particular interest in scholarly discourse since 2021. The paper includes a preliminary analysis of the literature followed by a series of recommendations for further research participation across the iFields, and plans for future work.

The study overall aims to begin bridging a gap between UAP Studies as an evolving research area and Library & Information Science (LIS), along with other closely related and established perspectives including Science of Science [12], Sociology of Science and Technology [13], and research data management studies [14]. This exploration of scholarly literature about UAP aligns with the broader goals of these fields to understand how scientific knowledge evolves and gains legitimacy over time through consensus. The paper also aims to contribute to ongoing conversations surrounding the nature and study of UAP, a previously stigmatized "fringe" topic that potentially represents an emerging interdisciplinary scientific field in need of novel information and data related research, support, and services.

## 2   Methods

To better understand the current landscape of scholarly literature on the topic of UAP, an exploratory scoping review was conducted following the PRISMA guidelines for systematic reviews [15], guided by the research question (RQ): "*How has the topic of UFOs/UAP evolved as a subject of scholarly inquiry over time?*". The following query was searched via Web of Science: ("ufo" OR "ufos" OR "uap" OR "uaps"). The query was limited to title and abstract, and results were further refined to include books, conference proceedings, and scholarly articles in English or Spanish. To ensure that Library & Information Science sources were sufficiently captured, the same search was conducted via ProQuest across three LIS databases: Library & Information Science Collection, Library & Information Science Abstracts (LISA), and Library Science Database. Most results were retrieved via Web of Science (n=1,999), with 219 results retrieved via the LIS databases. Duplicate records were processed (n=28), and the remaining records (n=2,190) were manually screened for relevance (criteria for relevance was considered to be sources directly addressing the topic of UFOs or UAP as a central theme). Thirty-seven potentially relevant items were indexed as citations only and unable to be retrieved. The final dataset includes 174 sources. Figure 1 shows the PRISMA flow diagram.

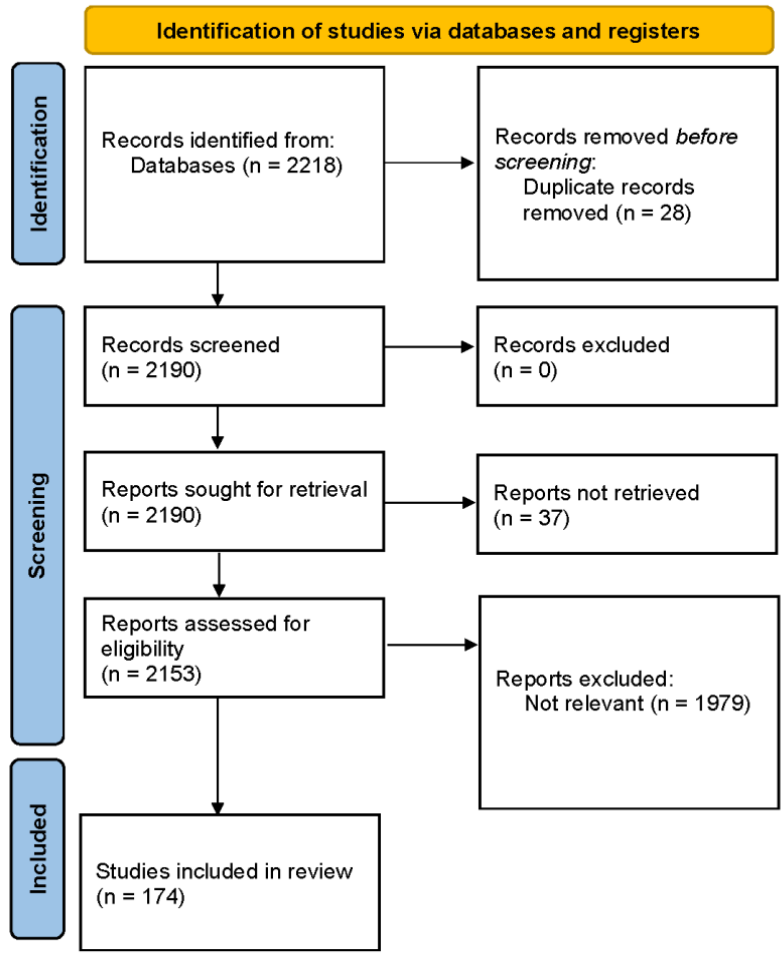

**Fig. 1.** PRISMA flow diagram of records retrieved and included.

For the purpose of this preliminary scoping review, a high-level content analysis was conducted to capture general features of the sources such as research area and year of publication. Items published since the 2021 release of the ODNI report [4] were examined more closely.

## 3  Results

Literature sources within the full dataset were published between 1967 and 2023 (Figure 2).

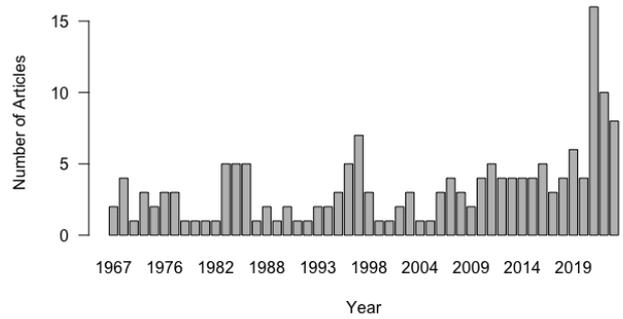

**Fig. 2.** Distribution of publication year.

Psychology (21%) and Religion (13%) are overwhelmingly dominant as research areas represented in the literature (per Web of Science classifications), followed by Astronomy & Astrophysics (6%), and Arts & Humanities (5%). See Figure 3.

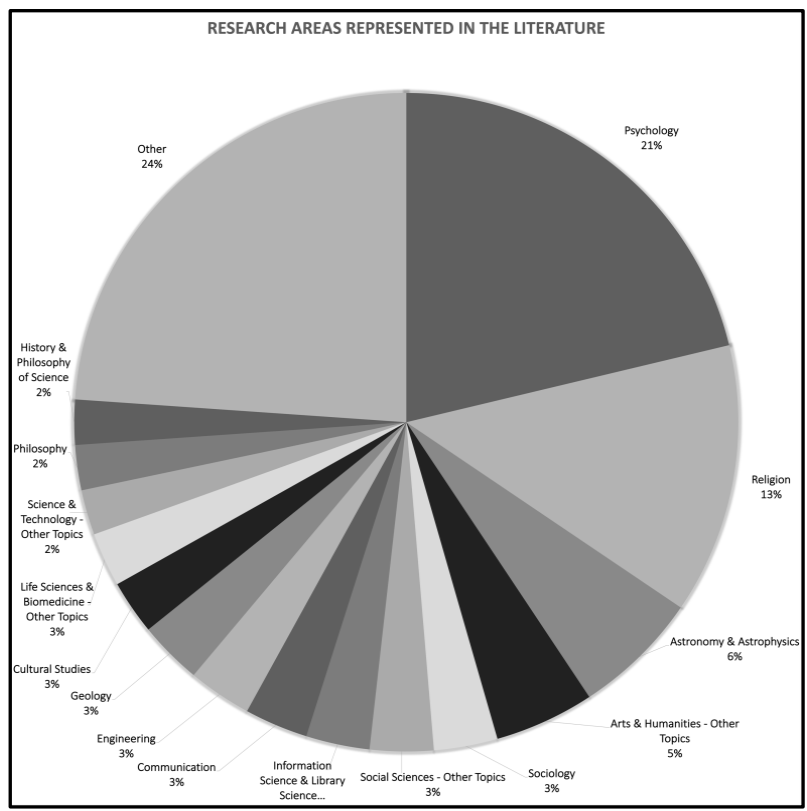

**Fig. 3.** Research areas represented in the literature (n=174).

For sources published since 2021 (n=34), Religion represents the largest proportion of research areas (30%), followed by Astronomy & Astrophysics (11%), Arts & Humanities (7%) and Communication (7%). However, apart from two items, the sources categorized under Religion (n=12) are part of a series [16], and therefore Astronomy & Astrophysics may be considered the true dominant research area. Note that Library & Information Science does not appear as a research area for these recent papers. The subset of 34 sources published since 2021 are categorized and summarized below.

### 3.1    Physical & Natural Sciences

Three papers describe methods and instruments for capturing UAP data. Two of these papers [17, 18] are affiliated with the Harvard-based Galileo Project that is searching for evidence of alien technology [19]; the former highlights motivations for studying UAP along with a roadmap for deploying equipment and implementing multi-sensor data processing pipelines, while the latter outlines computational techniques for detecting anomalous phenomena in data. The third instrumentation paper [20] presents a camera system design and software for calibrating the system to capture images of airborne objects for analysis of their movement patterns.

Antonio, et al. [21] analyze a large dataset of UFO reports, finding reporting patterns with respect to time of day, as well as increased reporting activity surrounding media attention. Wu & Yang [22] present a theoretical foundation for aircraft anti-gravity propulsion. Smith [23] speculates about the nature of extraterrestrial technology for interstellar probes, while Zuckerman [24] challenges a current hypothesis that an interstellar object that recently passed through the solar system was alien technology [25].

### 3.2    Social Sciences (including Communication, Psychology and Religious Studies)

In cultural studies, Fians [26] points to a need for anthropologists to take "native" perspectives seriously to avoid further marginalizing groups including UFO witnesses as Others. Marchena Sanabria [27] shows how narratives such as UFO reports were sensationalized by the media to distract from political corruption in Costa Rica between 1979 and 1985. Hayes [28] explores Cold War-era tensions between state actors and citizens surrounding UFO theories and narratives. Wright [29] takes a philosophical stance to examine the underexplored implications of alien and UFO tourism for the tourism industry.

Social science approaches include Yingling, et al.'s [3] large survey of faculty members showing that many academics think UAP is an important topic for research. Stise, et al. [30] analyze national survey data collected after the ODNI report release [4], showing an association between media use and belief in UFOs. Also following on the release of the ODNI report, Braum [31] shows that people form favorable opinions towards conspiracy theories in general when reputable politicians acknowledge that

UFOs may be extraterrestrial visitors. Adorjan & Kelly [32] leverage UFO-related "missing time" experiences to explore the importance and role of temporality in social constructionist scholarship. McVittie & McKinlay [33] investigate the discourse of news presenters speaking about UAP as they distance themselves from explaining the phenomena.

Religious studies scholars typically frame UFO and UAP as religious experiences. Agrama [34] challenges the effectiveness of secular science in light of recent developments in the study of UFOs. Similarly, Zeller [35] points to "enchanted" underpinnings of UFO investigations despite their secular organizations and approaches. Kivari [36] discusses how personal supernatural and paranormal experiences are integrated into broader social narratives. Finally, the Zeller [16] series *Handbook of UFO Religions* tackles a number of topics ranging from specific cultural case studies [37] to study of scholarship on UFOs and religion [38].

### 3.3 Arts & Humanities (including History and Philosophy)

Within the humanities-classified sources, Hodges & Paxton-Fear [39] analyze the writings that influenced Heaven's Gate cult members as the texts evolved from recruitment to reinforcing belief. Rose [40] explores racial aspects of UFO abduction narratives, suggesting that these stories may stem from a collective sense of white guilt regarding African enslavement and abduction accounts. Presenting a historical perspective, Guimont & Baumhammer [41] describe a panel series that debated the role of pseudoscience in the history and public understanding of science. Rooted in philosophical schools of thought, Butman [42] and Smith & Jonathan [43] explore the epistemology and miraculous nature of UFOs respectively.

## 4  Discussion

These results represent a snapshot of scholarly literature and discourse about UAP at the time of this writing (September 2023). Researchers from diverse areas are approaching UAP as a serious and actionable topic with implications for society and humanity (while stopping short of endorsing an extraterrestrial hypothesis). Circumstances are likely to change and evolve quickly in the future, although it is presently uncertain whether UAP Studies will develop into a recognized field. Nevertheless, credible research is underway and gaining publicity, suggesting a trend towards legitimizing UAP Studies as an interdisciplinary research area. To uphold the recommendations of the recent ODNI [4] and NASA [6] reports, a need remains to shift the focus from viewing UAP experiences as purely religious or psychological phenomena originating with the spirit or mind towards collecting standardized and reputable data across physical contexts and geographical locations, ranging from civilian and military reporting to high quality images and sensor data within and beyond Earth's atmosphere.

Notably, nearly all research areas and publications outlined in the Results section align with some aspect of the iFields' interdisciplinary research expertise, positioning these areas well for leadership in an emerging, multifaceted field of UAP Studies, with the potential to support government initiatives and inform policy development and knowledge construction. As uncovered by Yingling, et al. [3], academics are largely curious about this topic and consider it very important or essential to dispense with stigma and explore the nature of UAP using scientific methods regardless of the eventual explanations for the phenomena. A few ideas for iField participation in UAP Studies research are outlined below. Note that the broad categories of work presented in the following sections as headings are based upon the author's qualitative perspective and prior research and teaching experience, rather than on an existing taxonomy of subject areas.

### 4.1 Data Curation & Knowledge Organization

As emphasized by NASA [6], UAP-related data often lack standardization and may not be initially suited for analysis. Data curation and FAIR data practices are essential, particularly to support drawing conclusions from data collected across sources and contexts [7]. iFields are well positioned to help ensure that appropriate infrastructures are in place while assisting various disciplines and research communities with developing strategies to make data FAIR and ensuring that data are compatible with tools for analysis. Furthermore, data curation experts can contribute to handling distributed sensor-generated data, assessing data quality, and supporting anomaly detection techniques to identify UAP-related information that falls outside known constraints. Structured approaches to data management and information organization are crucial to ensure the integrity of a growing knowledge body of knowledge about UAP.

### 4.2 Information Behavior, Social Informatics, & Online Communities

Understanding information behavior and social dynamics surrounding UAP is another key, opportune area for research. Especially considering that UAP is a historically stigmatized topic and has been long associated with conspiracy theories, human information behavior and social informatics researchers can investigate how individuals seek, share, and evaluate information about UAP amidst a shifting narrative towards open curiosity and scientific study. Such research can shed light on the formation and evolution of online and offline communities and their impact on the dissemination of credible information about UAP as well as their belief systems. Additionally, studying various communities through a socio-technical lens can illuminate how they function and influence public perception, which is important for a comprehensive understanding of the phenomena and to support scientific investigation.

### 4.3 Data Science, Artificial Intelligence & Machine Learning

Data-driven approaches can be leveraged to make sense of UAP. For example, AI and ML techniques can be applied to analyze large amounts of data associated with UAP to identify patterns, trends, and anomalies that might not be apparent through traditional methods. Machine learning models can be trained with various data types, including sensor readings, eyewitness testimonies, and historical legacy data and records. Natural language processing (NLP) techniques can also extract valuable information from textual sources such as government reports and historical documents. As emphasized in the NASA UAPIST report [6], key challenges for researching and understanding UAP are data oriented.

**4.4  Library Services**

iFields can also contribute by exploring how libraries and other knowledge institutions should effectively collect, curate, and make UAP-related resources accessible to the public and researchers. This may include creating specialized collections, providing research assistance and reference support, and promoting critical thinking and information literacy skills related to UAP as a topic of growing interest to the public.

**4.5  Mis/dis-information**

The topic of UAP is especially vulnerable to misinterpretation and sensationalism, perhaps in part because it touches upon the existential question of whether humans are alone in the universe. As UAP Studies may become increasingly mainstream, ensuring communication of accurate and credible scientific information about the phenomena and in relation to previously established scientific movements such as the Search for Extraterrestrial Intelligence (SETI) is critical. iFields researchers can investigate various aspects of these challenges, including pinpointing sources and dissemination of misleading information as well as strategies for identifying and countering problematic narratives. This includes studying the role of media, online platforms, and information ecosystems in shaping public perception and belief systems surrounding UAP.

**4.6  Data & Science Literacy**

The topic of UAP may be an ideal entry point to engage the public in understanding and promoting data and science literacy. Educational programs and resources could be developed about UAP Studies to enhance the public's ability to critically assess and interpret data and scientific findings. This includes creating accessible materials that explain scientific methodologies and encourage evidence-based thinking, helping individuals to make informed judgments about UAP claims and research.

**4.7  Science & Technology Studies**

In the context of UAP, Science & Technology Studies researchers can explore the dynamics between scientific authorities, government agencies, and the public in defining and studying the phenomena. This includes analyzing the social construction of knowledge around UAP and how it intersects with government secrecy, military technology, international politics, and public perception. STS can also investigate the role of technological advancements in UAP observations, such as the impact of sophisticated sensors and data collection tools on the data available for analysis. By placing UAP within the broader framework of science and technology, it may be possible to develop a more comprehensive understanding of the cultural, political, and epistemological dimensions of UAP Studies and the phenomena itself.

## 5      Conclusion and Limitations

This paper has presented a preliminary literature review and set of recommendations for involvement of iFields in UAP Studies. The paper notes that LIS is underrepresented in the literature on this topic, indicating an opportunity to close a gap and apply demonstrated research strengths to understanding and explaining the societal, informational, and technological aspects of the phenomena. Future work by the author will build upon this initial exploration to address some of the iField research opportunities listed above and further analyze and expand a bibliometric dataset for in-depth qualitative analysis over the past 50+ years.

The present study has some limitations, by virtue of its early-stage nature. Primarily, as the study intentionally prioritized recall over precision, many irrelevant results were returned, which presented challenges for manually reviewing all sources with close attention to detail. Also, the query may have overlooked some relevant sources by focusing only on acronyms (UFO, UAP) instead of full phrases such as "Unidentified Flying Object(s)" and "Unidentified Anomalous Phenomena" (though the acronyms are typically included alongside the full phrases). Future work will also further explore and adjust the search strategy to capture relevant sources to the extent possible.

UAP Studies represents a rich and fascinating realm for further research and learning across the iFields. The NASA UAPIST report [6] that inspired this paper concludes, "there is an intellectual continuum between extrasolar technosignatures, solar system SETI, and potential unknown alien technology operating in Earth's atmosphere. If we recognize the plausibility of any of these, then we should recognize that all are at least plausible" (p. 33). The first two (SETI and the search for alien technosignatures) are represented by small but established research communities and are supported by instruments such as the James Webb Space Telescope [44]. As new communities, instruments and infrastructures take shape to support research and communication about UAP, this "intellectual continuum" provides a natural home for information studies work.